\documentclass[12pt]{iopart}

\usepackage{graphicx}
\begin{document}
\title[]{Super quantum discord for a class of two-qubit states with weak measurement}
\author{H. Eftekhari, E. Faizi}
\address{Physics Department, Azarbaijan shahid madani university 53714-161, Tabriz, Iran}
\ead{h.eftekhari@azaruniv.edu, efaizi@azaruniv.edu}
\begin{abstract}
So far, super quantum discord has been calculated explicitly only for Bell-diagonal states and expressions for more general quantum states are not known. In this paper, we derive explicit expressions for super quantum discord for a larger class of two-qubit states, namely, a 4-parameter family of two-qubit states.  We observe that, weak measurements obtain more quantumness of correlations than strong measurements. As an application, the dynamic behavior of the super quantum discord under decoherence channel is investigated. We find that, the super quantum discord decrease monotonically as a function of the measurement strength parameter.
\end{abstract}
{\bf Keywords:} Super quantum discord, Quantum discord, Two-qubit states, Dephasing channel.
\section{Introduction}
Weak measurement proposed by Aharonov, Albert, and Vaidman (AAV) \cite{Y. Aharonov} in 1988, is universal in the sense that any generalized measurement can be considered as a sequence of weak measurement which result in small changes to the quantum state for all outcomes \cite{O. Oreshkov}. weak measurement is very beneficial and helpful to understatnd many counterintuitive quantum phenomena such as Hardy's paradoxes \cite{Aharonov}. In the last years, much improvement have been done in this field, containing weak measurement involved in the contribution of probe dynamics \cite{A. Di Lorenzo}, like as weak measurement with ideal probe \cite{L. M. Johansen}, entangled probe \cite{D. Menzies}, and so on. Moreover, weak measurement deduced by some experiments is very helpful for measurements with high-precision. For instance, Hosten and Kwiat \cite{O. Hosten} utilize the weak measurement to study the spin Hall effect in light; Dixen et al. \cite{P. B. Dixon} use the weak measurement to indicate very small transverse beam deflections; Gillett et al. \cite{G. G. Gillett} apply the weak measurement to examine the feadback control of quantum systems with the existence of noise.\\
The quantum entanglement is a special kind of quantum correlation, but not the same with quantum correlation. It is accepted that the quantum correlations are more comprehensive than entanglement \cite{C. H. Bennett, W. H. Zurek}. Another measure of quantum correlation is the quantum discord \cite{H. Ollivier} which quantifies the quantumness of correlations in quantum states from a
measurement perspective. Up to now, numerous works have been made toward the significance and applications of quantum
discord. Particularly, there are few analytical expressions for quantum discord for two-qubit states, such as X states \cite{M. Ali, B. Li}.\\
Quantum discord is a quantum correlation based on von Neumann measurement. Because of the essential role of weak measurement, it is interesting to know how quantum discord will be with weak measurement? Lately, it is shown that weak measurement done on one
of the subsystems can lead to "super quantum discord" that is always greater than the normal quantum discord captured by the projective (strong) measurements \cite{U. Singh}. We want to know whether weak measurements
can always obtain more quantumness of correlations than normal quantum discord for a bipartite quantum systems? If they
can, then one can exploit this extra quantum correlation for
information processing. In this article, we evaluate explicit expressions for super quantum discord for a class of two-qubit states, namely, a 4-parameter family of two-qubit states in Sec. 2. In Sec. 3 the dynamic behavior of super quantum discord under decoherence is investigated. A brief conclusion is given in Sec. 4.
\section{The super quantum discord for a class of X-states}
The quantum discord for a bipartite quantum state $\rho_{AB}$ with the projective measurement $\{\Pi_i^B\}$ done on the subsystem $B$ is defined as the difference between the total correlation $I(\rho_{AB})$ \cite{M. H. Partovi} and classical correlation $J(\rho_{AB})$ \cite{L. Henderson}, that is,
\begin{eqnarray}D(\rho_{AB})=\min_{\{\Pi_i^B\}}\Sigma_ip_iS(\rho_{A|i})+S(\rho_B)-S(\rho_{AB})
\end{eqnarray}
with the minimization is to be done over all possible projection-valued measurements $\{\Pi_i^B\}$, where $S(\rho)=-\tr(\rho\log_2\rho)$ is the von Neumann entropy, and $\rho_B$ is the reduced density matrix for the part $B$ and
\begin{eqnarray}p_i=\tr_{AB}[(I_A\otimes{\Pi_i^B})\rho_{AB}(I_A\otimes{\Pi_i^B})],\\\nonumber
\rho_{A|i}=\frac{1}{p_i}\tr_B[(I_A\otimes{\Pi_i^B})\rho_{AB}(I_A\otimes{\Pi_i^B})].
\end{eqnarray}
The weak measurement operators are given by \cite{O. Oreshkov}
\begin{eqnarray}P(x)=\sqrt{\frac{(1-\tanh{x})}{2}}\Pi_0+\sqrt{\frac{(1+\tanh{x})}{2}}\Pi_1,\\\nonumber
P(-x)=\sqrt{\frac{(1+\tanh{x})}{2}}\Pi_0+\sqrt{\frac{(1-\tanh{x})}{2}}\Pi_1,
\end{eqnarray}
where $x$ indicates the measurement strength parameter, $\Pi_0$ and $\Pi_1$ are two orthogonal projectors and $\Pi_0+\Pi_1=I$. The weak measurement operators satisfy: (i) $P^\dagger(x)P(x)+P^\dagger(-x)P(-x)=I$, (ii) $\lim_{x\rightarrow{\infty}}P(x)=\Pi_0$ and $\lim_{x\rightarrow{\infty}}P(-x)=\Pi_1$.\\
Lately, Singh and Pati introduce the super quantum discord of any bipartite quantum state $\rho_{AB}$ with weak measurement on the subsystem $B$ \cite{U. Singh}, the super quantum discord specified by $D_w(\rho_{AB})$ is given by
\begin{eqnarray}D_w(\rho_{AB})=\min_{\{\Pi_i^B\}}S_w(A|\{{P^B(x)}\})+S(\rho_B)-S(\rho_{AB})
\end{eqnarray}
with the minimization is to be done over all possible projection-valued measurements $\{\Pi_i^B\}$, where $S(\rho)=-\tr(\rho\log_2\rho)$ is the von Neumann entropy of a quantum state $\rho_{AB}$, $\rho_B$ is the reduced density matrix of $\rho_{AB}$ for the subsystem $B$, and
\begin{eqnarray}S_w(A|\{P^B(x)\})=p(x)S(\rho_A|\{P^B(x)\})+p(-x)S(\rho_A|\{P^B(-x)\}),
\end{eqnarray}
 \begin{eqnarray}p(\pm(x))=\tr_{AB}[(I\otimes{P^B(\pm{x})})\rho_{AB}(I\otimes{P^B(\pm{x})})],
\end{eqnarray}
\begin{eqnarray}\rho_{A|P^B(\pm{x})}=\frac{\tr_{B}[(I\otimes{P^B(\pm{x})})\rho_{AB}(I\otimes{P^B(\pm{x})})]}{\tr_{AB}[(I\otimes{P^B(\pm{x})})\rho_{AB}(I\otimes{P^B(\pm{x})})]},
\end{eqnarray}
$\{P^B(x)\}$ is weak measurement operators carried out on the subsystem B.\\
So far super quantum discord has been calculated explicitly only for Bell-diagonal states \cite{Y. K. Wang}. The great difficulty is that we can not even able to find the value of super quantum discord for the 5-parameter family of X-states. In this paper, we will calculate the super quantum discord for the full 4-parameter family of X-states with additional assumptions. We consider the following 4-parameter quantum system
\begin{eqnarray}
\rho_{AB}=\frac{1}{4}\left(
\begin{array}{cccccccc}
1+s+c_3&&0&&0&&c_1-c_2\\
0&&1-s-c_3&&c_1+c_2&&0\\
0&&c_1+c_2&&1+s-c_3&&0\\
c_1-c_2&&0&&0&&1-s+c_3\\
\end{array}
\right).
\end{eqnarray}
we will only cosider the following simplified family of Eq.(8), where
\begin{eqnarray}|c_1|<|c_2|<|c_3|, 0<|s|<1-|c_3|
\end{eqnarray}
 The eigenvalues of the state in Eq.(8) are given by
\begin{eqnarray} \lambda_{1,2}=\frac{1}{4}[1-c_3\pm{\sqrt{s^2+(c_1+c_2)^2}}],\\\nonumber
\lambda_{3,4}=\frac{1}{4}[1+c_3\pm{\sqrt{s^2+(c_1-c_2)^2}}].
\end{eqnarray}
The entropy $\rho_{AB}$ is given by
\begin{eqnarray}S(\rho_{AB})=-\Sigma_{i=1}^4\lambda_i\log_2{\lambda_i}\\\nonumber
=2-\frac{1}{4}[(1-c_3+\sqrt{s^2+(c_1+c_2)^2})\log_2(1-c_3+\sqrt{s^2+(c_1+c_2)^2})\\\nonumber
+(1-c_3-\sqrt{s^2+(c_1+c_2)^2})\log_2(1-c_3-\sqrt{s^2+(c_1+c_2)^2})\\\nonumber
+(1+c_3+\sqrt{s^2+(c_1-c_2)^2})\log_2(1+c_3+\sqrt{s^2+(c_1-c_2)^2})\\\nonumber
+(1+c_3-\sqrt{s^2+(c_1-c_2)^2})\log_2(1+c_3-\sqrt{s^2+(c_1-c_2)^2})].
\end{eqnarray}
 Let $\{\Pi_k=|k\rangle\langle{k}|, k=0,1\}$, be the local measurement for the subsystem $B$ along the computational base $|k\rangle$. Then any weak measurement operators for the subsystem $B$ can be given as \cite{Y. K. Wang}:
\begin{eqnarray}I\otimes{P(\pm{x})}=\sqrt{\frac{(1\mp{\tanh{x}})}{2}}I\otimes{V\Pi_0V^\dagger}+\sqrt{\frac{(1\pm{\tanh{x}})}{2}}I\otimes{V\Pi_1V^\dagger}
\end{eqnarray}
for some unitary $V\in{{U}}(2)$. We may write any $V\in{U}(2)$ as
\begin{eqnarray}V=tI+i\vec{y}\vec{\sigma}
\end{eqnarray}
 with $t\in{R}$, $\vec{y}=(y_1,y_2,y_3)\in{R^3}$ and $t^2+y_1^2+y_2^2+y_3^2=1$.
After the weak measurement, the state $\rho_{AB}$ will turn to the ensemble $\{\rho_{A|P^B(\pm(x))},p(\pm{x})\}$. We need to calculate $\rho_{A|P^B(\pm{x})}$ and $p(\pm{x})$. We use the relations in Ref.\cite{S. Luo},
\begin{eqnarray}{V^\dagger}\sigma_1{V}=(t^2+y_1^2-y_2^2-y_3^2)\sigma_1+2(ty_3+y_1y_2)\sigma_2+2(-ty_2+y_1y_3)\sigma_3,\\\nonumber
{V^\dagger}\sigma_2{V}=2(-ty_3+y_1y_2)\sigma_1+(t^2+y_2^2-y_1^2-y_3^2)\sigma_2+2(ty_1+y_2y_3)\sigma_3,\\\nonumber
 {V^\dagger}\sigma_3{V}=2(ty_2+y_1y_3)\sigma_1+2(-ty_1+y_2y_3)\sigma_2+(t^2+y_3^2-y_1^2-y_2^2)\sigma_3,
\end{eqnarray}
and $\Pi_0\sigma_3\Pi_0=\Pi_0, \Pi_1\sigma_3\Pi_1=-\Pi_1, \Pi_j\sigma_k\Pi_j=0, for j=0,1, k=1,2$, from Eqs.(6) and (7), we find $p(\pm{x})=\frac{1\mp{sz_3{\tanh{x}}}}{2}$ and
\begin{eqnarray}\rho_{A|P^B(+x)}=\frac{1}{2(1-sz_3\tanh{x})}\times\\\nonumber
(I-\tanh{x}(sz_3I+c_1z_1\sigma_1+c_2z_2\sigma_2+c_3z_3\sigma_3)),\\\nonumber
\rho_{A|P^B(-x)}=\frac{1}{2(1+sz_3\tanh{x})}\times\\\nonumber
(I+\tanh{x}(sz_3I+c_1z_1\sigma_1+c_2z_2\sigma_2+c_3z_3\sigma_3)),
\end{eqnarray}
where $z_1=2(-ty_2+y_1y_3), z_2=2(ty_1+y_2y_3), z_3=t^2+y_3^2-y_1^2-y_2^2$.\\
To simplify we write $X=sz_3I+c_1z_1\sigma_1+c_2z_2\sigma_2+c_3z_3\sigma_3$ and Eq. (15) can be modify to
\begin{eqnarray}\rho_{A|P^B(+x)}=\frac{1}{2(1-sz_3\tanh{x})}(I-X\tanh{x}),\\\nonumber
\rho_{A|P^B(-x)}=\frac{1}{2(1+sz_3\tanh{x})}(I+X\tanh{x}).
\end{eqnarray}
The eigenvalues of $\frac{1}{2(1-sz_3\tanh{x})}(I-X\tanh{x})$ and $\frac{1}{2(1+sz_3\tanh{x})}(I+X\tanh{x})$ are $\lambda_5=\frac{1+(\phi+\theta)\tanh{x}}{2(1-\phi\tanh{x})}$, $\lambda_6=\frac{1+(\phi-\theta)\tanh{x}}{2(1-\phi\tanh{x})}$ and $\lambda_7=\frac{1+(-\phi-\theta)\tanh{x}}{2(1+\phi\tanh{x})}$, $\lambda_8=\frac{1+(-\phi+\theta)\tanh{x}}{2(1+\phi\tanh{x})}$ respectively, where $\theta$ and $\phi$ are as follows:
\begin{eqnarray}\phi=sz_3,\theta=\sqrt{|c_1z_1|^2+|c_2z_2|^2+|c_3z_3|^2}.
\end{eqnarray}
Therefore
\begin{eqnarray}S(\rho_{A|P^B(+x)})=-\frac{1+(\phi+\theta)\tanh{x}}{2(1-\phi\tanh{x})}\log_2{\frac{1+(\phi+\theta)\tanh{x}}{2(1-\phi\tanh{x})}}\\\nonumber
-\frac{1-(\phi-\theta)\tanh{x}}{2(1-\phi\tanh{x})}\log_2{\frac{1-(\phi-\theta)\tanh{x}}{2(1-\phi\tanh{x})}}
\end{eqnarray}
and
\begin{eqnarray}S(\rho_{A|P^B(-x)})=-\frac{1+(-\phi-\theta)\tanh{x}}{2(1+\phi\tanh{x})}\log_2{\frac{1+(-\phi-\theta)\tanh{x}}{2(1+\phi\tanh{x})}}\\\nonumber
-\frac{1+(-\phi+\theta)\tanh{x}}{2(1+\phi\tanh{x})}\log_2{\frac{1+(-\phi+\theta)\tanh{x}}{2(1+\phi\tanh{x})}}
\end{eqnarray}
thus form Eq.(5) we have
\begin{eqnarray}S_w({A|\{P^B(x)\}})=f(\theta,\phi)=\frac{(1-\phi\tanh{x})}{2}S(\rho_{A|P^B(+x)})\\\nonumber
+\frac{(1+\phi\tanh{x})}{2}S(\rho_{A|P^B(-x)})\\\nonumber
=-\frac{1+(\phi+\theta)\tanh{x}}{4}\log_2{\frac{1+(\phi+\theta)\tanh{x}}{2(1-\phi\tanh{x})}}\\\nonumber
-\frac{1+(\phi-\theta)\tanh{x}}{4}\log_2{\frac{1+(\phi-\theta)\tanh{x}}{2(1-\phi\tanh{x})}}\\\nonumber
-\frac{1+(-\phi-\theta)\tanh{x}}{4}\log_2{\frac{1+(-\phi-\theta)\tanh{x}}{2(1+\phi\tanh{x})}}\\\nonumber
-\frac{1+(-\phi+\theta)\tanh{x}}{4}\log_2{\frac{1+(-\phi+\theta)\tanh{x}}{2(1+\phi\tanh{x})}}.
\end{eqnarray}
By using of the domain of logarithmic function in $f(\theta,\phi)$ and Eq.(9), we can find the range of $\theta$ and $\phi$:
\begin{eqnarray}0\leq{|c_1|}\leq{\theta}\leq{|c_3|}\leq{1}, -1<{\phi}<1.
\end{eqnarray}
one can see that $f(-\phi,\theta)=f(\phi,\theta)$, and $f(\phi,\theta)$ is symmetric with respect to the $\theta$; $\frac{\partial{f}}{\partial{\theta}}=-\frac{1}{4}\log[\frac{((1+\theta{tanh{x}})^2-\phi^2tanh^2{x})(1+\phi\tanh{x})^2}{((1-\theta{tanh{x}})^2-\phi^2tanh^2{x})(1-\phi\tanh{x})^2}]<0$, $0<{\theta}<1$, $f(\phi,\theta)$ is a function which decreasing monotonous; $\frac{\partial{f}}{\partial{\phi}}=-\frac{1}{4}\log[\frac{((1+{\phi}tanh{x})^2-\theta^2tanh^2{x})(1+\phi\tanh{x})^2}{((1-{\phi}tanh{x})^2-\theta^2tanh^2{x})(1-\phi\tanh{x})^2}]<0$, $0<{\phi}<1$, $f(\phi,\theta)$ is a function which decreasing monotonous. When $\theta=|c_3|$ by $z_1^2+z_2^2+z_3^2=1$, Eqs.(9) and (17) we obtain
\begin{eqnarray}\phi=|s|.
\end{eqnarray}
By using of Eq.(9) the projection of $f(\phi,\theta)$ on the plane $\phi{o}\theta$ is a symmetric rectangle with respect to the $\theta$-axis and by applying of the monotonicity of $f(\phi,\theta)$ in the positive direction of $\theta$ and $\phi$, we can obtain the minimum of $f(\phi,\theta)$ at the point $(|s|,|c_3|)$. Therefore the minimum of $S_w(A|\{P^B(x)\})$ is as follows:
\begin{eqnarray}\min{S_w(A|\{P^B(x)\})}=
-{\frac{(1+(s+c_3)\tanh{x})}{4}}\log_2{\frac{(1+(s+c_3)\tanh{x})}{2(1-s\tanh{x})}}\\\nonumber
-{\frac{(1+(s-c_3)\tanh{x})}{4}}\log_2{\frac{(1+(s-c_3)\tanh{x})}{2(1-s\tanh{x})}}\\\nonumber
-{\frac{(1+(-s-c_3)\tanh{x})}{4}}\log_2{\frac{(1+(-s-c_3)\tanh{x})}{2(1+s\tanh{x})}}\\\nonumber
-{\frac{(1+(-s+c_3)\tanh{x})}{4}}\log_2{\frac{(1+(-s+c_3)\tanh{x})}{2(1+s\tanh{x})}}.
\end{eqnarray}
Then, by Eqs.(4), (11) and $S(\rho_B)=1-\frac{1}{2}[(1-s)\log_2(1-s)+(1+s)\log_2(1+s)]$, the super quantum discord of the state in Eqs.(8),(9) is given by
\begin{eqnarray}D_w(\rho^{AB})=-\frac{1}{4}[{(1+(s+c_3)\tanh{x})}\log_2{\frac{(1+(s+c_3)\tanh{x})}{2(1-s\tanh{x})}}\\\nonumber
+{(1+(s-c_3)\tanh{x})}\log_2{\frac{(1+(s-c_3)\tanh{x})}{2(1-s\tanh{x})}}\\\nonumber
{(1+(-s-c_3)\tanh{x})}\log_2{\frac{(1+(-s-c_3)\tanh{x})}{2(1+s\tanh{x})}}\\\nonumber
{(1+(-s+c_3)\tanh{x})}\log_2{\frac{(1+(-s+c_3)\tanh{x})}{2(1+s\tanh{x})}}]\\\nonumber
-\frac{1}{2}[(1-s)\log_2(1-s)+(1+s)\log_2(1+s)]\\\nonumber
+\frac{1}{4}[(1-c_3+\sqrt{s^2+(c_1+c_2)^2})\log_2(1-c_3+\sqrt{s^2+(c_1+c_2)^2})\\\nonumber
+(1-c_3-\sqrt{s^2+(c_1+c_2)^2})\log_2(1-c_3-\sqrt{s^2+(c_1+c_2)^2})\\\nonumber
+(1+c_3+\sqrt{s^2+(c_1-c_2)^2})\log_2(1+c_3+\sqrt{s^2+(c_1-c_2)^2})\\\nonumber
+(1+c_3-\sqrt{s^2+(c_1-c_2)^2})\log_2(1+c_3-\sqrt{s^2+(c_1-c_2)^2})].
\end{eqnarray}
The quantum discord of state (8) is given by (see Ref.\cite{B. Li})
\begin{eqnarray}Q(\rho^{AB})=1+f(s)+\Sigma_{i=1}^4\lambda_i\log_2{\lambda_i}+min\{S_1,S_2,S_3\}
\end{eqnarray}
where $S_1$, $S_2$ and $S_3$ are given by:
\begin{eqnarray}S_1=-\frac{1+s+c_3}{4}\log_2{\frac{1+s+c_3}{2(1+s)}}-\frac{1+s-c_3}{4}\log_2{\frac{1+s-c_3}{2(1+s)}}\\\nonumber
-\frac{1-s-c_3}{4}\log_2{\frac{1-s-c_3}{2(1-s)}}
-\frac{1-s+c_3}{4}\log_2{\frac{1-s+c_3}{2(1-s)}}
\end{eqnarray}
\begin{eqnarray}S_2=1+f(c_1)
\end{eqnarray}
\begin{eqnarray}S_3=1+f(c_2)
\end{eqnarray}
and
\begin{eqnarray}f(x)=-\frac{1-x}{2}\log_2(1-x)-\frac{1+x}{2}\log_2(1+x)
\end{eqnarray}

\begin{figure}
\includegraphics[width=2.8in]{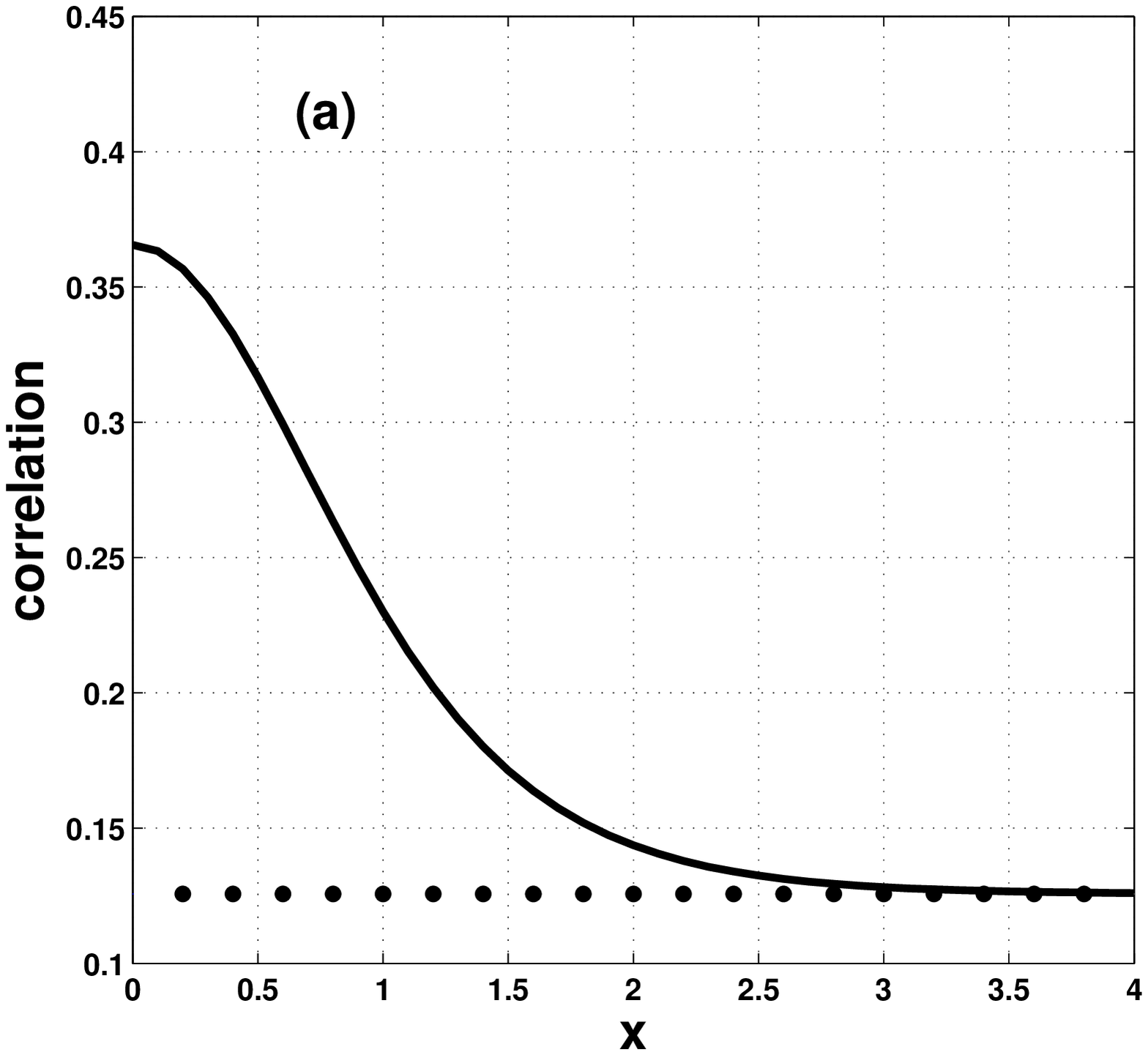}
\includegraphics[width=2.8in]{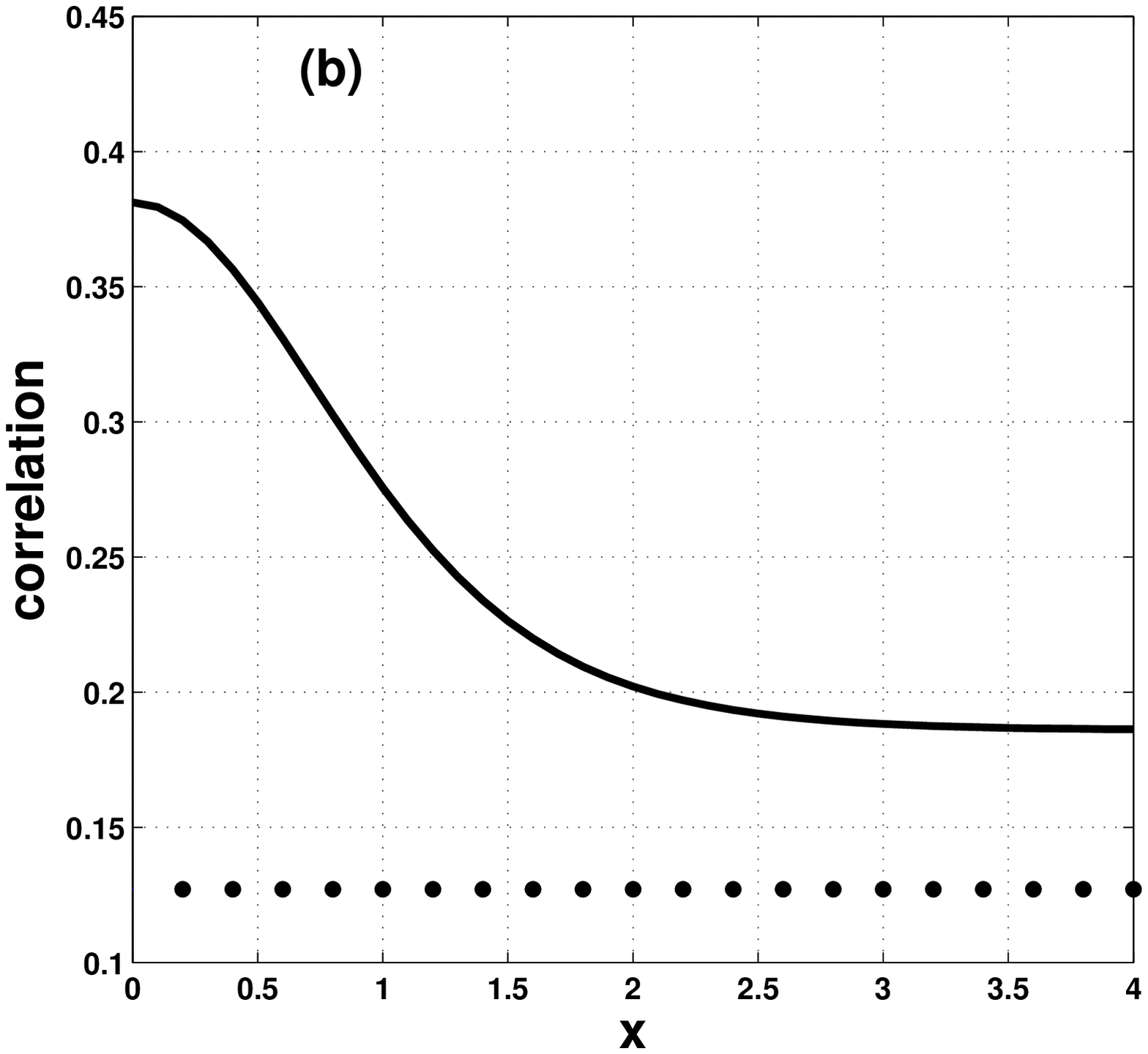}
\caption{Super quantum discord (solid line) and quantum discord (dotted line) for $c_1=0.3$, $c_2=-0.4$ and $c_3=0.56$ and $s=0$ (a) $s=0.2$ (b)}
 \label{fig1}
\end{figure}
We plot the super quantum discord and the quantum discord with respect to $x$ in Fig.1. In Fig.1(a) we take $c_1=0.3$, $c_2=-0.4$, $c_3=0.56$ and $s=0$ (we recall that in this case our state reduce to Bell-diagonal state), we can see that, at first the super quantum discord is greater than the normal discord for smaller values of x and approaches to the normal discord for larger values of $x$. In Fig.1(b) we take $s=0.2$ and the other parameters are same as Fig.1(a). It is noticeable that, the super quantum discord is greater than the normal discord.
\section{Dynamics of super quantum correlation under local nondissipative channels}
Here we investigate the effect of phase flip channel on the states in Eqs.(8), (9)\cite{J. Maziero}. The Kraus operators for phase flip channel are given by: $\Gamma_0^{(A)}=\mathrm{diag}(\sqrt{1-p/2},\sqrt{1-p/2})\otimes{I}$, $\Gamma_1^{(A)}=\mathrm{diag}(\sqrt{p/2},-\sqrt{p/2})\otimes{I}$, $\Gamma_0^{(B)}={I}\otimes{\mathrm{diag}}{(\sqrt{1-p/2},\sqrt{1-p/2})}$, $\Gamma_1^{(B)}={I}\otimes{\mathrm{diag}}{(\sqrt{p/2},-\sqrt{p/2})}$, where $p=1-\exp(-\gamma{t})$, in which $\gamma$ indicates the phase damping rate \cite{J. Maziero, T. Yu}. \\
We use $\varepsilon(.)$ as the operator of decooherence. Then under the phase flip channel, we have
\begin{eqnarray}\varepsilon(\rho)=\frac{1}{4}(I\otimes{I}+I\otimes{s\sigma_3}+(1-p)^2c_1\sigma_1\otimes{\sigma_1}\\\nonumber
+(1-p)^2c_2\sigma_2\otimes{\sigma_2}+c_3\sigma_3\otimes{\sigma_3}).
\end{eqnarray}
The super quantum discord of the state Eq.(8) under phase flip channel is given by
\begin{eqnarray}ND_w(\rho^{AB})=-\frac{1}{4}[{(1+(s+c_3)\tanh{x})}\log_2{\frac{(1+(s+c_3)\tanh{x})}{2(1-s\tanh{x})}}\\\nonumber
+{(1+(s-c_3)\tanh{x})}\log_2{\frac{(1+(s-c_3)\tanh{x})}{{2(1-s\tanh{x})}}}\\\nonumber
+{(1+(-s-c_3)\tanh{x})}\log_2{\frac{(1+(-s-c_3)\tanh{x})}{{2(1+s\tanh{x})}}}\\\nonumber
+{(1+(-s+c_3)\tanh{x})}\log_2{\frac{(1+(-s+c_3)\tanh{x})}{{2(1+s\tanh{x})}}}]\\\nonumber
-\frac{1}{2}[(1-s)\log_2(1-s)+(1+s)\log_2(1+s)]\\\nonumber
+\frac{1}{4}[(1-c_3+\sqrt{s^2+(1-p)^4(c_1+c_2)^2})\log_2(1-c_3+\sqrt{s^2+(1-p)^4(c_1+c_2)^2})\\\nonumber
+(1-c_3-\sqrt{s^2+(1-p)^4(c_1+c_2)^2})\log_2(1-c_3-\sqrt{s^2+(1-p)^4(c_1+c_2)^2})\\\nonumber
+(1+c_3+\sqrt{s^2+(1-p)^4(c_1-c_2)^2})\log_2(1+c_3+\sqrt{s^2+(1-p)^4(c_1-c_2)^2})\\\nonumber
+(1+c_3-\sqrt{s^2+(1-p)^4(c_1-c_2)^2})\log_2(1+c_3-\sqrt{s^2+(1-p)^4(c_1-c_2)^2})].
\end{eqnarray}
\begin{figure}
\includegraphics[width=2.8in]{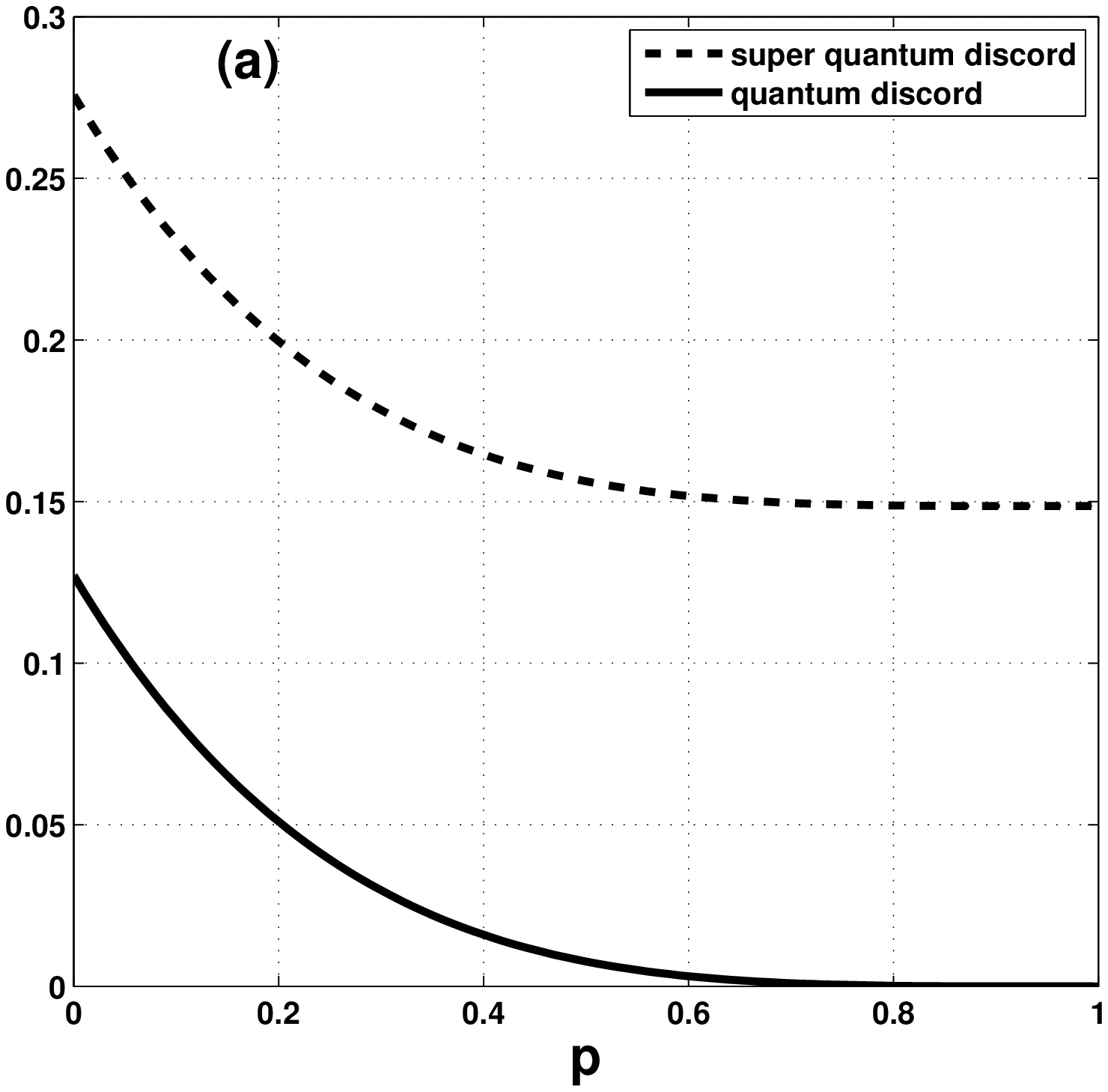}
\includegraphics[width=2.8in]{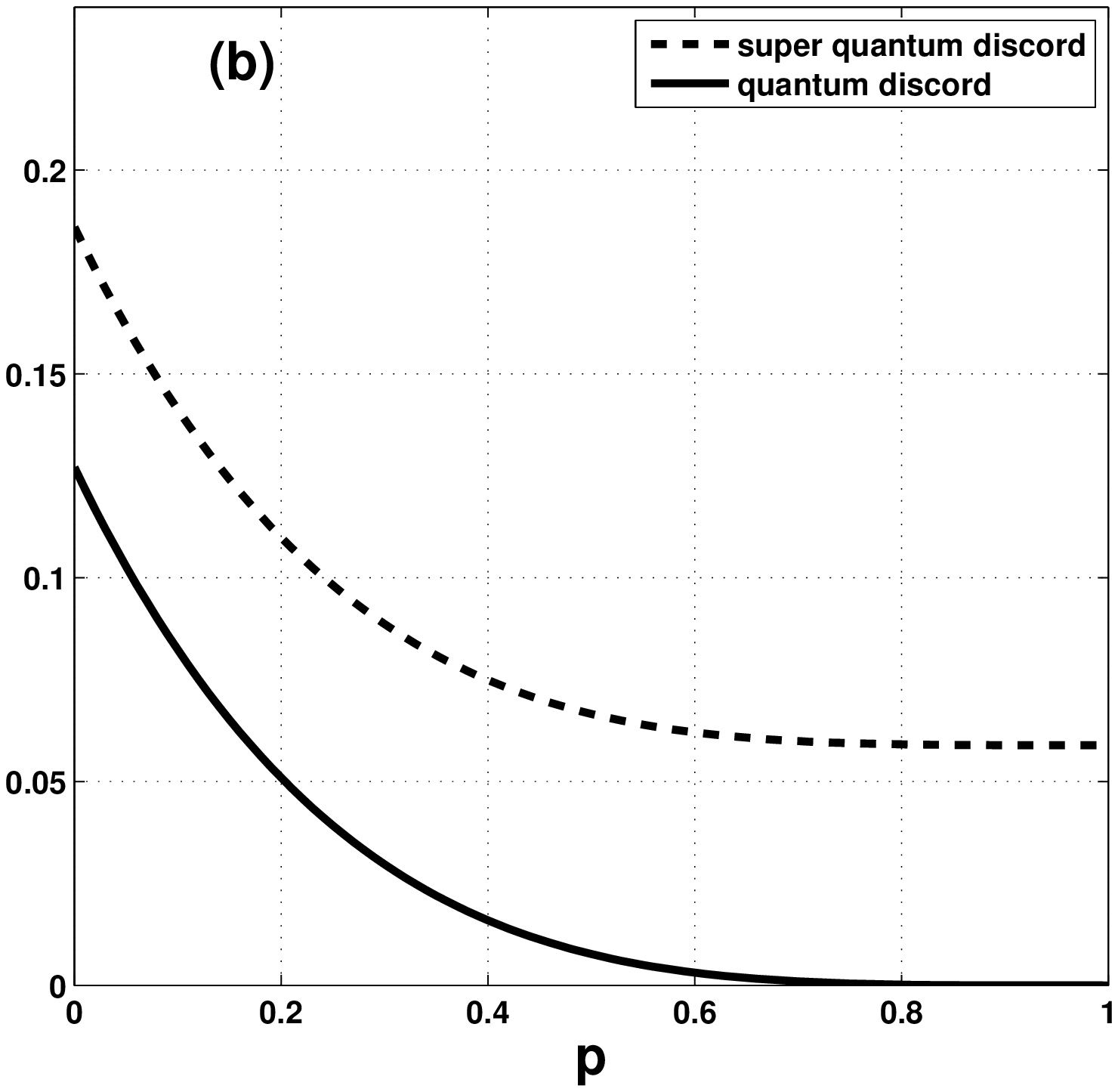}
\includegraphics[width=3.3in]{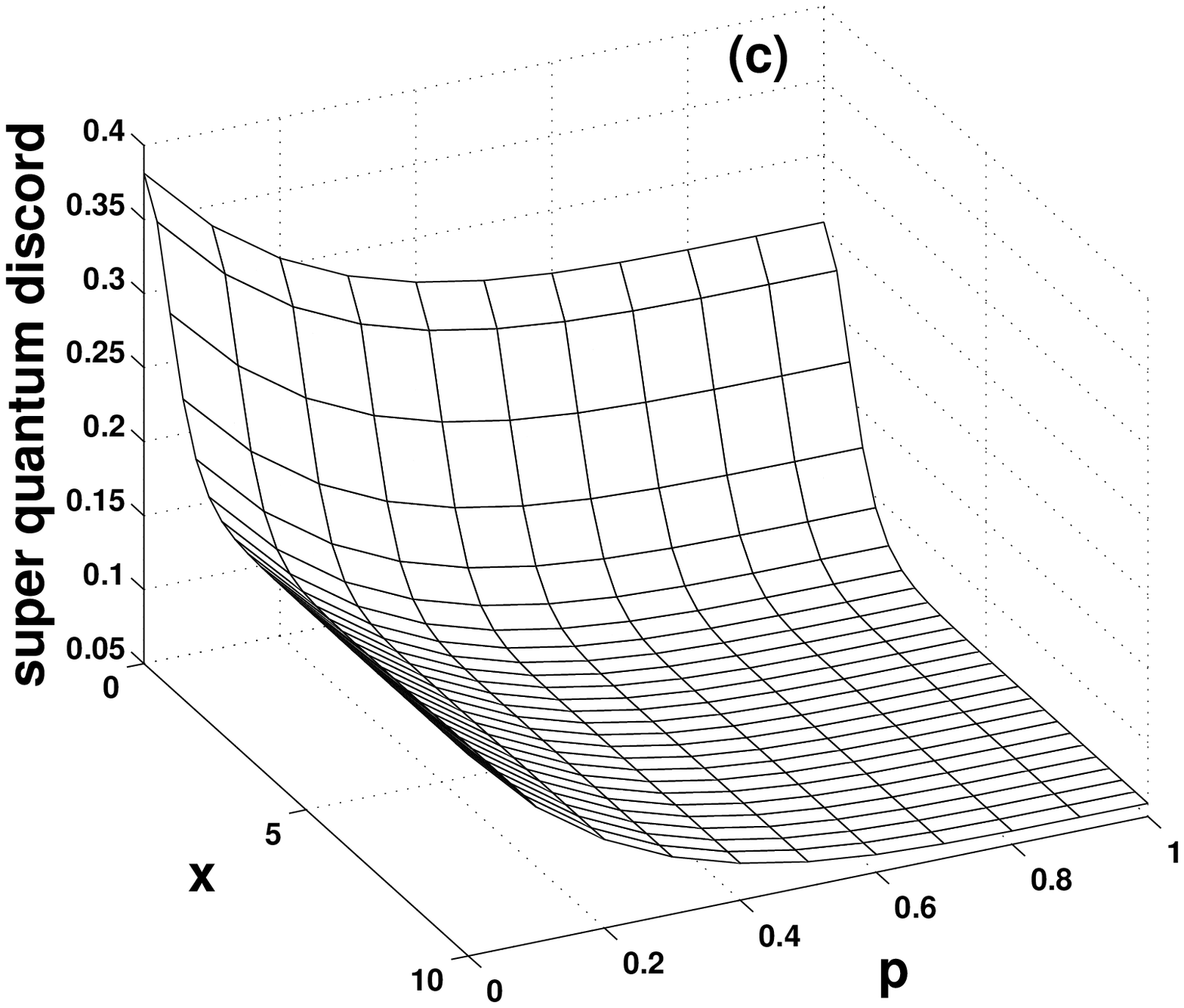}
\caption{Super quantum discord and quantum discord under the phase flip channel: super quantum discord (dashed line) and quantum discord (solid line) as a function of $p$ for $c_1=0.3$, $c_2=-0.4$ and $c_3=0.56$ and $s=0.2$ (a) $x=1$ (b) $x=5$ and (c) super quantum discord as a function of $x$ and $p$ for $c_1=0.3$, $c_2=-0.4$ and $c_3=0.56$ and $s=0.2$}
 \label{fig1}
\end{figure}
We plot in Fig.2 the super quantum discord and quantum discord for the state in Eq.(8) and (9) under phase flip channel. Again, we take $c_1=0.3$, $c_2=-0.4$, $c_3=0.56$ and $s=0.2$. Fig. 2(a) and 2(b) show the behavior of super quantum discord and quantum discord versus $p$ for $x=1$ and $x=5$ respectively. We find that super quantum discord is greater than quantum discord for $x=1$ (Fig.2(a)) and $x=5$ (Fig.2(b)). The super quantum discord under the phase flip channel as a function of $x$ and $p$ is shown in (c); we can see that the super quantum discord decreases by increasing x and p.
\section{Conclusion}
To conclude, in this work the super quantum discord has been calculated analytically for a 4-parameter family of X-states with additional assumptions. It is noticeable that, weak measurement induced quantum discord, called as the "super quantum discord" is larger than the normal quantum discord captured by the strong measurement. Therefore, the notion of super quantum discord can be a useful resource for quantum information processing tasks, quantum communication and quantum computation. Moreover, the dynamics of super quantum discord for phase flipping channel has been studied. The results indicate that super quantum discord decreases by increasing $x$ and $p$.
\section{Acknowledgement}
This research has been supported by Azarbaijan Shahid Madani university.


\section{References}
\begin{thebibliography}{0}
\bibitem{Y. Aharonov}Y. Aharonov, D. Z. Albert, and L. Vaidman, {\em Phys. Rev. Lett.} {\bf 60}, 1351 (1988).
\bibitem{O. Oreshkov}O. Oreshkov and T. A. Brun, {\em Phys. Rev. Lett.} {\bf 95}, 110409 (2005).
\bibitem{Aharonov}Y. Aharonov, A. Botero, S. Pospescu, B. Reznik, and J. Tollaksen, {\em Phys. Lett. A} {\bf 301}, 130 (2002); J. S. Lundeen
and A. M. Steinberg, {\em Phys. Rev. Lett.} {\bf 102}, 020404 (2009); K. Yokota, T. Yamamoto, M. Koashi, and N. Imoto,
{\em New J. Phys.} {\bf 11}, 033011 (2009).
\bibitem{A. Di Lorenzo}A. Di Lorenzo and J. C. Egues, {\em Phys. Rev. A} {\bf 77}, 042108 (2008).
\bibitem{L. M. Johansen}L. M. Johansen, Phys. Rev. Lett. 93, 120402 (2004).
\bibitem{D. Menzies}D. Menzies and N. Korolkova, Phys. Rev. A 77, 062105 (2008).
\bibitem{O. Hosten}O. Hosten and P. Kwiat, {\em Science} {\bf 319}, 787 (2008); K. J. Resch, {\em ibid.} {\bf 319}, 733 (2008).
\bibitem{P. B. Dixon}P. B. Dixon, D. J. Starling, A. N. Jordan, and J. C. Howell, {\em Phys. Rev. Lett.} {\bf 102}, 173601 (2009); J. C. Howell,
D. J. Starling, P. B. Dixon, P. K. Vudyasetu, and A. N. Jordan, {\em Phys. Rev. A} {\bf 81}, 033813 (2010).
\bibitem{G. G. Gillett}G. G. Gillett, R. B. Dalton, B. P. Lanyon, M. P. Almeida, M. Barbieri, G. J. Pryde, J. L. O'Brien, K. J. Resch,
S. D. Bartlett, and A. G. White, {\em Phys. Rev. Lett.} {\bf 104}, 080503 (2010).
\bibitem{C. H. Bennett}C. H. Bennett, D. P. Di Vincenzo, C. A. Fuchs, T. Mor, E. Rains, P. W. Shor, J. A. Smolin and W. K. Wootters,
{\em Phys. Rev. A} {\bf 59}, 1070 (1999).
\bibitem{W. H. Zurek}W. H. Zurek, {\em Ann. Phys. (Leipzip)} {\bf 9}, 5 (2000); L. Henderson and V. Vedral, {\em J. Phys. A} {\bf 34}, 6899 (2001).
\bibitem{H. Ollivier}H. Ollivier and W. H. Zurek, {\em Phys. Rev. Lett.} {\bf 88}, 017901 (2001).
\bibitem{M. Ali}M. Ali, A. R. P. Rau, G. Alber, {\em Phys. Rev. A} {\bf 81}, 042105 (2010).
\bibitem{B. Li}B. Li, Z. X. Wang and S. M. Fei, {\em Phys. Rev. A} {\bf 83}, 022321 (2011); Q. Chen, C. Zhang, S. Yu, X. X. Yi and C. H. Oh, {\em Phys. Rev. A} {\bf 84}, 042313 (2011); M. Shi, C. Sun, F. Jiang, X. Yan and J. Du, {\em Phys. Rev. A} {\bf 85}, 064104 (2012); S. Vinjanampathy, A. R. P. Rau, {\em J. Phys. A} {\bf 45}, 095303 (2012).
 \bibitem{U. Singh}U. Singh and A. K. Pati, arX iv:1211.0939.
\bibitem{M. H. Partovi}M. H. Partovi, {\em Phys. Lett. A} {\bf 137}, 445 (1989).
\bibitem{L. Henderson}L. Henderson and V. Vedral, {\em J. Phys. A} {\bf 34}, 6899 (2001).
\bibitem{Y. K. Wang}Y. K. Wang, T. Ma, H. Fan, S. M. Fei, Z. X. Wang, {\em quantum information processing} {\bf 13}, 283 (2014).
\bibitem{S. Luo}S. Luo, Phys. Rev. A 77, 042303 (2008).
\bibitem{B. Li}B. Li, Z. X. Wang, S. M. Fei, {\em phys. Rev. A} {\bf 88}, 022321 (2011).
\bibitem{J. Maziero}J. Maziero, L. C. Celeri, R. M. Serra and V. Vedral, Phys. Rev. A 80, 044102 (2009).
\bibitem{T. Yu}T. Yu and J. H. Eberly, Phys. Rev. Lett. 97, 140403 (2006).
\end{thebibliography}
\end{document}